\documentclass[a4paper,11pt]{article}

\usepackage{jcappub} 

\usepackage[T1]{fontenc} 
\usepackage{tensor}
\usepackage{physics}
\usepackage{subcaption} 
\title{\boldmath Static stable timelike circular orbits and Aschenbach effect in horizonless solutions of Einsteinian cubic gravity}

\author[a,1]{Zhen-Hua Zhao,\note{Corresponding author.}}
\author[b,2]{Yong-Qiang Wang,\note{Corresponding author.}}

\affiliation[a]{Department of Applied Physics, Shandong University of Science and Technology, Qingdao 266590, China}
\affiliation[b]{Lanzhou Center for Theoretical Physics, School of Physical Science and Technology, Lanzhou University, Lanzhou 730000, China}

\emailAdd{zhaozhh78@sdust.edu.cn}
\emailAdd{yqwang@lzu.edu.cn}

\abstract{
In modified gravity theories, horizonless compact objects serve as a compelling alternative to black holes for testing strong-field gravity. Einsteinian cubic gravity (ECG) provides a gravitational framework for constructing these viable astrophysical models.  
We investigate the existence, stability, and observable signatures of static stable timelike circular orbits (SSTCOs) in static spherically symmetric ECG horizonless spacetimes.  
We derive timelike geodesic equations, construct the effective potential for circular orbits, and perform a numerical integration to verify orbital stability.   
We confirm that SSTCOs exist in both solution branches of ECG horizonless objects and that their radii coincide with the innermost stable circular orbit (ISCO). The Aschenbach effect manifests as a non-monotonic radial dependence of a zero angular momentum observer (ZAMO) measured velocity. Furthermore, we find that the stability of circular orbits exhibits a 'double stable region' structure. As the specific energy \(E\) of a test particle transitions from the outer edge to the inner edge (i.e., the ISCO) of the inner stable region, its variation can exceed \(1\) (i.e., \(\Delta E > 1\)), implying that during this process, the gravitational system can release an amount of energy exceeding the rest mass of the particle itself.

}

\begin{document}
\maketitle
\flushbottom

\section{Introduction}

The study of black holes and their horizonless alternatives provides a theoretical foundation for testing general relativity and modified gravity theories. 
In addition to gravitational-wave detection~\cite{AbbottAbbottAbbott2016}, observing their optical images through surrounding accretion matter provides another crucial approach~\cite{AkiyamaAlberdiAlef2019e,AkiyamaAlberdiAlef2022e}, since these objects themselves do not emit electromagnetic radiation directly.
This requires  a thorough understanding of the motion of accreting particles and photons, as special particle orbits can produce unique observable signatures. 

Recently, Collodel, Kleihaus, and Kunz discovered a novel class of orbits around rotating compact objects (e.g., boson stars, hairy black holes, and wormholes) where massive particles remain at rest relative to a distant static observer \cite{CollodelKleihausKunz2018}. These static timelike circular orbits arise from the interplay between the particle's angular momentum and the frame-dragging effect induced by spacetime rotation. Subsequent studies have shown that similar static torus structures may exist around rotating boson stars \cite{TeodoroCollodelKunz2021} and scalarized Kerr black holes \cite{TeodoroCollodelDoneva2021}. Beyond rotating spacetimes, static orbits have also been found in charged, static spherically symmetric black hole solutions \cite{WeiZhangLiu2023} and in massive Einstein-Maxwell gravity \cite{YerraMukherjiBhamidipati2025}, forming what are termed static spheres.

The existence of static timelike circular orbits implies a characteristic region: within it, the angular velocity of massive particles, as seen by a distant observer, decreases as the orbital radius decreases. This behavior is interpreted as the \textit{Aschenbach effect} in static spherically symmetric spacetimes \cite{WeiLiu2023,YerraMukherjiBhamidipati2025,AfsharSadeghi2025}. Originally discovered in the context of rapidly rotating Kerr black holes (with spin parameter $a > 0.9953M$), the Aschenbach effect describes a non-monotonic variation in the orbital velocity $v^{(\phi)}$ measured by a ZAMO for test particles on circular orbits \cite{Aschenbach2004}. Aschenbach proposed that this non-monotonic velocity profile could excite a distinct 3:1 resonance between radial and vertical epicyclic frequencies, potentially providing a means to estimate black hole masses \cite{Aschenbach2004a,Aschenbach2006}. The effect has since been investigated in various contexts, including topology changes of von Zeipel surfaces \cite{StuchlikSlanyTorok2005}, rapidly rotating Kerr-(A)dS black holes \cite{MuellerAschenbach2007}, braneworld Kerr black holes \cite{StuchlikBlaschkeSlany2011}, charged particles around weakly magnetized rotating black holes \cite{TursunovStuchlikKolos2016}, and spinning particles in Kerr and Kerr-(A)dS spacetimes \cite{KhodagholizadehPerlickVahedi2020,VahediKhodagholizadehTursunov2021}.

However, all known examples of static timelike circular orbits rely on either spacetime rotation, electric charge, or additional matter fields. Whether such orbits can exist in a purely gravitational, electrically neutral, and  static setting remains an open question.

A necessary condition for static timelike circular orbits in static spherically symmetric black hole solutions is that the metric component $g_{tt}$ possesses at least one extremum. This suggests that similar orbits might exist in horizonless solutions where $g_{tt}$ satisfies this condition. Recently, horizonless solutions \cite{Wang2024a} in ECG \cite{BuenoCano2016} have been reported. In these solutions $g_{tt}$ typically exhibits extrema, which makes static orbits possible. Unlike previous models requiring additional matter fields or charges, ECG provides a purely gravitational framework, which offers a cleaner setting for studying these phenomena.

In this work, we investigate horizonless solutions in ECG and demonstrate the existence of SSTCOs for massive particles in a purely gravitational, electrically neutral, static, and spherically symmetric context. We further identify these orbits as the ISCOs. Additionally, we investigate the radial distributions of the velocity, angular velocity, and specific energy for particles on circular orbits, particularly in connection with the Aschenbach effect.

The paper is organized as follows: Section \ref{model} introduces Einsteinian cubic gravity and its numerical solutions. Section \ref{orbits} presents the geodesic equations and analyzes the existence and stability of static orbits. Section \ref{aschenbach} discusses the Aschenbach effect in these spacetimes. Finally, Section \ref{conclusions} summarizes our results and discusses their implications.

\section{Einsteinian cubic gravity  (ECG)}\label{model}

ECG is a higher-order gravitational theory constructed from cubic powers of the Riemann tensor. It has attracted considerable attention due to its unique properties in various aspects. 
The theory not only aligns with general relativity in its linear perturbation spectrum, ensuring the propagation of massless gravitons without ghost fields, but also possesses dimension-independent coefficients in its cubic curvature term.
Notably, ECG is neither trivial nor topological in four-dimensional spacetime, thereby providing a nontrivial effective model for third-order corrections to gravitational theory. Recently, there has been extensive research on ECG models and their extensions. For example, in the context of black holes, studies have covered spherically symmetric solutions \cite{BuenoCano2016a,HennigarMann2017}, generalizations to rotating and charged cases \cite{BurgerEmondMoynihan2020,AdairBuenoCano2020,CanoPereniguez2020,FrassinoRocha2020,KordZangenehKazemi2020,SajadiHendi2022,LessaSilva2023}, and black string solutions \cite{Bakhtiarizadeh2022,BakhtiarizadehGolchin2022}. Furthermore, ECG has been extended to research in quasi-topological gravity \cite{BuenoCanoMoreno2019,CisternaGrandiOliva2020}, cosmology \cite{EricesPapantonopoulosSaridakis2019,QuirosGarcia-SalcedoGonzalez2020,CisternaGrandiOliva2020,Dengiz2025}, and wormholes \cite{MehdizadehZiaie2019,HussainMustafa2022,MustafaXiaHussain2020,MustafaAtamurotovGhosh2023,LuYangMann2025}, demonstrating its broad potential for application across various fields of gravitational physics.

The action of four dimensional  ECG is given by \cite{BuenoCano2016a}:
\begin{equation}\label{action}
  S_G = \frac{1}{16 \pi G} \int \sqrt{-g} \, \dd^4 x \left( R - G^2 \lambda \mathcal{P} \right),
\end{equation}
where $G$ is the Newton gravitational constant, $R$ is the Ricci scalar, $\lambda$ is  a dimensionless gravitational coupling, and
\begin{align}
\mathcal{P} = 12 \tensor{R}{_a ^c _b ^d} \tensor{R}{_c ^e _d ^f} \tensor{R}{_e ^a _f ^b} + \tensor{R}{_{ab}^{cd}} \tensor{R}{_{cd}^{ef}} \tensor{R}{_{ef}^{ab}} \nonumber  \\
- 12 R_{abcd} R^{ac} R^{bd} 
+ 8 \tensor{R}{_a^b} \tensor{R}{_b ^c} \tensor{R}{_c ^a}.
\end{align}

For a static spherically symmetric spacetime, we adopt the line element:
\begin{equation}\label{metric}
\dd  s^2 = -f(r) \dd  t^2 + \frac{\dd  r^2}{f(r)} + r^2 \left( \dd \theta^2 + \sin^2 \theta \, \dd \phi^2 \right),
\end{equation}
where $f(r)$ satisfies the ordinary differential equation \cite{BuenoCano2016a}:
\begin{align}\label{eq4}
&-(f-1)r - G^2 \lambda \bigg[ 4f'^3 + 12 \frac{f'^2}{r} - 24 f (f-1) \frac{f'}{r^2} \nonumber \\
&- 12 f f'' \left( f' - \frac{2(f-1)}{r} \right) \bigg] = \mathcal{C},
\end{align}
where $\mathcal{C} = 2 G M$ and $M$ denotes the mass of the object.

Equation \eqref{eq4} is highly nonlinear, making it extremely challenging to obtain analytical solutions. Therefore, seeking asymptotic solutions near certain points \cite{BuenoCano2016a,HennigarMann2017} or complete numerical solutions \cite{Wang2024a} is a more practical approach to analyzing the problem.

It is important to note that although Eq. \eqref{eq4} appears as a second-order differential equation, it originates from the third-order field equations of ECG. Specifically, Eq. \eqref{eq4} represents a first integral of the original third-order equations, with the integration constant $\mathcal{C}$. Consequently, three conditions are required to uniquely determine the solution. The spacetime is asymptotically flat, which imposes the conditions:
\begin{equation}\label{bc1}
f(\infty) = 1 \quad \text{and} \quad f'(\infty) = 0.
\end{equation}
Here, $f(\infty)=1$ normalizes the time coordinate at infinity, while $f'(\infty)=0$ ensures true asymptotic flatness and determines the mass parameter $\mathcal{C}$ through the asymptotic expansion  \cite{BuenoCano2016} $f(r) \approx 1 - \mathcal{C}/r + \cdots$. Additionally, for horizonless solutions, we require regularity at the origin $r=0$. This provides the third condition:
\begin{equation}\label{bc2}
f(0) = f_0 > 0.
\end{equation}
The parameter $f_0 = f(0)$ specifies the central value of the metric function and must be strictly positive to avoid a curvature singularity.

Further, to simplify numerical computations, we introduce the following dimensionless variable substitutions:
\begin{equation}
\frac{r}{G M} \to r, \; \frac{\lambda}{G^2 M^4} \to \lambda, \; \frac{t}{G M} \to t, \; \text{and} \; \frac{\dd s}{G M} \to \dd s.
\end{equation}
After this re-scaling, $\lambda$ remains a dimensionless free parameter. Subsequently, equation \eqref{eq4} is simplified by setting
\begin{equation}
G \to 1, \; M \to 1,
\end{equation}
which leaves the redefined $\lambda$ as the only free parameter and $\mathcal{C}=2$.

To ensure the solution covers the point at infinity, we introduce a compactified coordinate $x = r/(1+r)$ to map the physical domain $r \in [0, \infty)$ onto the computational interval $x \in [0,1]$. The boundary value problem defined by Eq.~\eqref{eq4} with conditions \eqref{bc1} and \eqref{bc2} was solved using the finite element software FEniCSx \cite{AlnaesLoggOlgaard2014,ScroggsBarattaRichardson2022,BarattaDeanDokken2023}.
We set the computational domain to be discretized with $2 \times 10^4$ mesh nodes, and the relative numerical error tolerance is set to $10^{-8}$.

\begin{figure*}[htbp]
    \centering
    \subcaptionbox{}{%
        \includegraphics[width=0.45\textwidth]{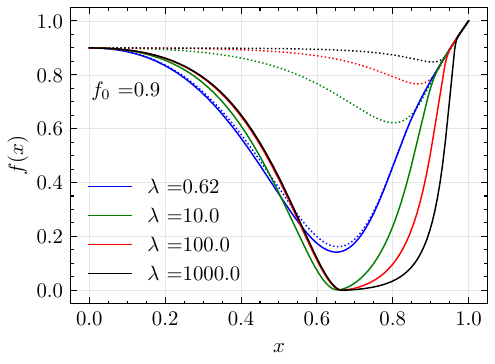}%
    }
    \subcaptionbox{}{%
        \includegraphics[width=0.455\textwidth]{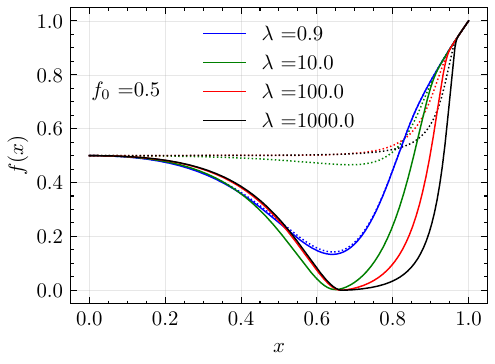}%
    }

 \subcaptionbox{}{%
        \includegraphics[width=0.45\textwidth]{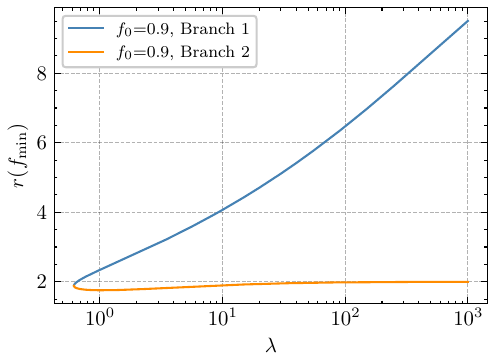}%
    }  
    \subcaptionbox{}{%
        \includegraphics[width=0.455\textwidth]{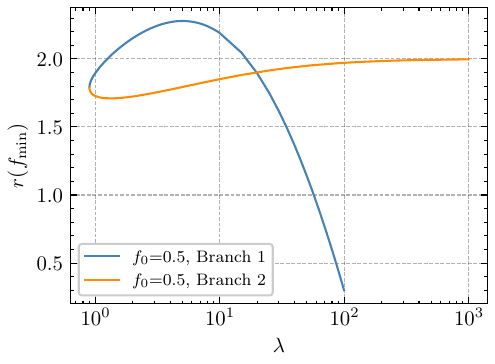}%
    }
\caption{
Solutions of $f(r)$ for (a) $f_0=0.9$ and (b) $f_0=0.5$. Dashed and solid lines of the same color correspond to Branch 1 and Branch 2, respectively. Panels (c) and (d) show the location of the local minimum of $f(r)$ as a function of $\lambda$ for $f_0=0.9$ and $f_0=0.5$, respectively. }
\label{sol1}
\end{figure*}

For given \(f_0\) and \(\lambda\), Eq.~\eqref{eq4} admits two independent branches of solutions, which we label as Branch 1 and Branch 2. Fig.~\ref{sol1} shows example solutions for \(f_0 = 0.9, 0.5\), where for a given \(\lambda\), dashed and solid lines of the same color correspond to the solutions in two branches, respectively. It can be seen from Fig.~\ref{sol1}a and ~\ref{sol1}b that the solutions from the two branches gradually converge as \(\lambda\) decreases.

Within the parameter range considered, not every solution of \( f(r) \) possesses a local minimum, but each solution contains at most one. The dependence of the location of this local minimum on the parameter \( \lambda \) is shown in Figs.~\ref{sol1}c and \ref{sol1}d. From the figures, we observe that: (1) As \( \lambda \) decreases, the locations of the local minima in the two branches approach each other; (2) For the Branch 1 solution with \( f_0 = 0.9 \), the location of the local minimum increases monotonically with \( \lambda \); for the Branch 1 solution with \( f_0 = 0.5 \), however, it first increases and then decreases, eventually disappearing when \( \lambda \ge 200 \); (3) In Branch 2, the location of the local minimum first decreases and then increases as \( \lambda \) grows.

\section{Static stable timelike circular orbits (SSTCOs)}\label{orbits}

The action of a unit-mass particle moving in this spacetime is given by
\begin{equation}\label{action-p}
S = \int g_{\mu\nu} \dot{x}^{\mu} \dot{x}^{\nu} \dd \tau.
\end{equation}
The dot denotes the derivative with respect to proper time $\tau$. 

Assuming the particle moves in the equatorial plane with $\theta = \pi/2$, the equations of motion are
\begin{align}
&\ddot{r} =\frac{1}{2 g_{rr}} \left( g_{tt,r} \dot{t}^2 - g_{rr,r} \dot{r}^2 + g_{\phi\phi,r} \dot{\phi}^2 \right), \label{eqr} \\ 
&\dot{\phi} = \frac{L}{g_{\phi\phi}}, \label{eqphi} \\
&\dot{t} = -\frac{E}{g_{tt}}, \label{tdot1}
\end{align}
where the subscript ``$,r$'' denotes the derivative with respect to the radial coordinate $r$. The conserved quantities $E$ and $L$ represent the energy and angular momentum per unit mass of the particle (specific  energy and specific  angular momentum), respectively.

Additionally, the particle’s motion satisfies the normalization condition:
\begin{equation}\label{geodesic}
g_{\mu\nu} \dot{x}^{\mu} \dot{x}^{\nu} = -1, 
\end{equation}
which can be reduced to 
\begin{equation}\label{geodesic1}
\frac{1}{2} \dot{r}^2 + V(r) = 0, 
\end{equation}
with 
\begin{equation}\label{V}
V(r) = \frac{1}{2} \frac{1}{g_{tt} g_{rr}} \left( E^2 + g_{tt} \left( 1 + \frac{L^2}{g_{\phi\phi}} \right) \right) = -\frac{1}{2}  E^2 + \frac{1}{2} f(r) \left( 1 + \frac{L^2}{r^2} \right) .
\end{equation}

For a timelike circular orbit (TCO) satisfying $\dot{r} = 0$, we have $
V(r_\mathrm{TCO}) = 0$, where $r_\mathrm{TCO}$ denotes the radius of the circular orbit. To analyze its stability, the following effective potential is introduced:
\begin{equation}\label{Veff}
V_\mathrm{eff}(r) = \frac{1}{2} f(r) \left( 1 + \frac{L^2}{r^2} \right).
\end{equation}

 When a massive particle moves in a circular orbit, its angular velocity is:
\begin{equation}\label{eqO}
\Omega_\mathrm{TCO} = \frac{\dd \phi}{\dd t} = \frac{\dot{\phi}}{\dot{t}} =\pm \sqrt{ \frac{ - \partial_r g_{tt} }{ \partial_r g_{\varphi\varphi} } } =\pm \sqrt{ \frac{f'(r) }{2 r}},
\end{equation}
where the prime denotes the derivative with respect to  $r$ and $f'(r)\ge 0$.

Moreover, the specific angular momentum  and the specific energy for a circular orbit are expressed as: 
\begin{equation}\label{L}
L_\mathrm{TCO}= \frac{g_{\varphi \varphi} \Omega_\mathrm{TCO}}{\sqrt{  D(r)}}, \ \ 
E_\mathrm{TCO} = - \frac{g_{tt}}{\sqrt{ D(r)}},
\end{equation}
where function
\begin{equation}\label{D}
 D(r) = -g_{tt} - g_{\varphi\varphi} \Omega_\mathrm{TCO}^2 = \frac{2f(r)- r f'(r)}{2}\ge 0.
\end{equation}

\begin{figure*}[htbp]
\centering

\subcaptionbox{Branch 1}{
\includegraphics[width=0.45\linewidth]{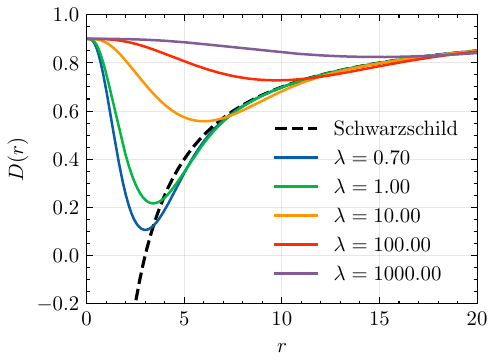}}
\subcaptionbox{Branch 2}{
\includegraphics[width=0.45\linewidth]{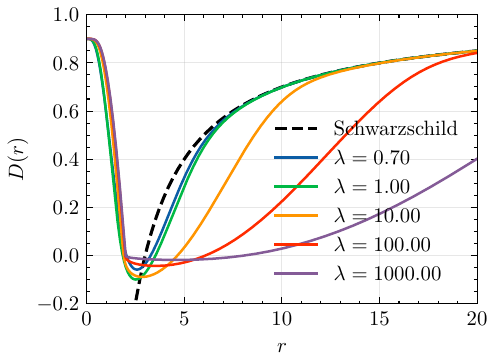}}

\caption{Plots of function $D(r)$ for different values of $\lambda$. The left panel (a) corresponds to Branch 1, while the right panel (b) corresponds to Branch 2. The dashed line represents the Schwarzschild case, and the solid lines represent the ECG case with parameters $f_0 = 0.9$ and different  values of $\lambda$.} 
\label{solD}
\end{figure*}

The function $D(r)$ for the two branches with $f_0 = 0.9$ is shown in Fig.~\ref{solD}. It can be observed that in Branch 2, the minimum value of $D(r)$ is less than zero, which is distinctly different from the case in Branch 1. Physically,  $D(r)<0$ implies that the denominator in $E$ and $L$ becomes imaginary, so no physical circular orbit exists in that region and the region allowing circular orbits becomes discontinuous.

\begin{figure}[htb]
\centering
\includegraphics[width=0.5\linewidth]{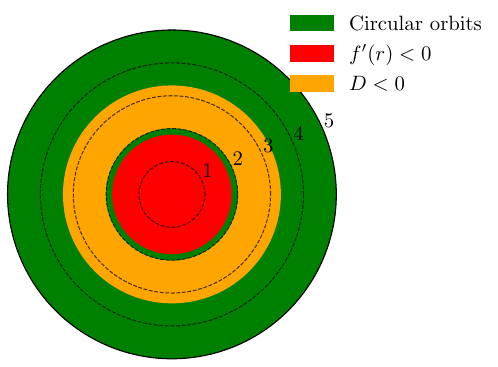}
\caption{The region where circular orbits exist (green). The corresponding solution comes from Branch 2, with parameters \( f_0 = 0.9 \) and \( \lambda = 0.7 \). Here, red region: f'(r)<0; orange region: D(r)<0; green region: allowed circular orbits. The black dashed circle serves as a reference, and the numbers indicate the values of the radius.} 
\label{solDf}
\end{figure}

If we further consider the condition $f'(r) \ge 0$, the range allowing for the existence of circular orbits in Branch 2 will be further constrained. As shown in Fig.~\ref{solDf}, we have marked the regions where $f'(r) < 0$ in red, the regions where $D(r) < 0$ in orange, and the regions that simultaneously satisfy $f'(r) \ge 0$ and $D(r) \ge 0$ (i.e., the regions where circular orbits are allowed) in green, with the parameters  $f_0 = 0.9$ and $\lambda = 0.7$. It can be observed that, under the premise of satisfying all physical conditions ($f'(r) \ge 0$ and $D(r) \ge 0$), the region allowing for the existence of circular orbits is divided into two discontinuous parts.

The necessary conditions for stable timelike circular orbits (STCOs) are:
\begin{equation}\label{cond}
\quad V_\mathrm{eff}'(r_\mathrm{STCO}) = 0 , \  V_\mathrm{eff}''(r_\mathrm{STCO}) \ge 0,
\end{equation}
where $r_\mathrm{STCO}$ denotes the radius of stable circular orbits  and $V_\mathrm{eff}''(r_\mathrm{STCO})= 0$ is the critical stability condition.

Let the radius of the SSTCO be \( r_{\mathrm{SSTCO}} \). On this orbit, the angular velocity of the particle satisfies \( \Omega(r_{\mathrm{SSTCO}}) = 0 \). From equations \eqref{eqO} and \eqref{L}, we obtain  
\begin{equation}  
 f'(r_{\mathrm{SSTCO}})  = 0, \ L_{\mathrm{TCO}}(r_{\mathrm{SSTCO}}) = 0  
\end{equation}  
respectively.
Substituting \(L_{\mathrm{TCO}}(r_{\mathrm{SSTCO}}) = 0\) and  $f'(r_{\mathrm{SSTCO}})  = 0$  into equation \eqref{Veff}, we have  
\begin{equation}  
 V_{\mathrm{eff}}'(r_{\mathrm{SSTCO}})  =  \frac{1}{2}f'(r_{\mathrm{SSTCO}}), \  
 V_{\mathrm{eff}}''(r_{\mathrm{SSTCO}}) =  \frac{1}{2}f''(r_{\mathrm{SSTCO}}).  
\end{equation}

In summary, the conditions for the existence of a SSTCO can be stated as  
\begin{equation}\label{conds3}
f(r_{\mathrm{SSTCO}}) > 0, \ 
 f'(r_{\mathrm{SSTCO}})  = 0, \
 f''(r_{\mathrm{SSTCO}})  \ge 0.  
\end{equation}

\begin{figure}[htb]
\centering
\includegraphics[width=0.5\linewidth]{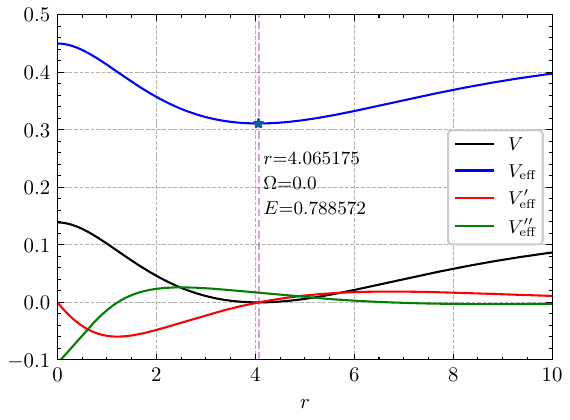}  
\caption{
Effective potential for particles with zero angular velocity $\Omega = 0$. The corresponding solution comes from Branch 1, with parameters $f_0 = 0.9$ and $\lambda = 10$. The blue star ($\star$) marks the location of the minimum of the effective potential. The vertical red dashed line indicates the location of the minimum of effective potential.}
\label{fVeff}
\end{figure}

As an example, Fig.~\ref{fVeff} shows the effective potential energy \(V_{\text{eff}}(r)\) and its first and second derivatives for particles with angular velocity \(\Omega = 0\) in Branch 1, with parameters \(f_0 = 0.9\) and \(\lambda = 10\). The figure reveals that the effective potential energy has a minimum at \(r = 4.065175\), where the conditions for a SSTCO are satisfied: \(V_{\text{eff}}'(r) = 0\) and \(V_{\text{eff}}''(r) > 0\). Thus, the radius of the corresponding SSTCO is determined to be \(r_\mathrm{SSTCO} = 4.065175\). Further substituting this radius into the condition \(V(r_{\mathrm{SSTCO}}) = 0\) yields the corresponding specific energy as \(E = 0.788572\).

Next, we substitute equation \eqref{tdot1} into equation \eqref{eqr} and proceed to numerically integrate equations \eqref{eqr}--\eqref{eqphi} to visually verify the stability of the aforementioned orbit. The verification is conducted in two steps.

\begin{figure*}[htb]
\centering
\subcaptionbox{}{
\includegraphics[width=0.45\linewidth]{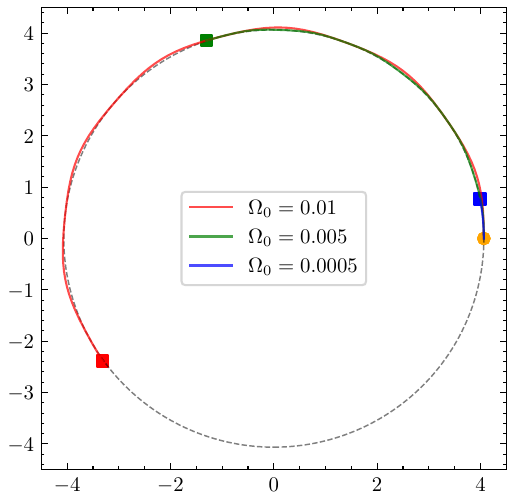}\label{fig63}
}
\subcaptionbox{}{
\includegraphics[width=0.45\linewidth]{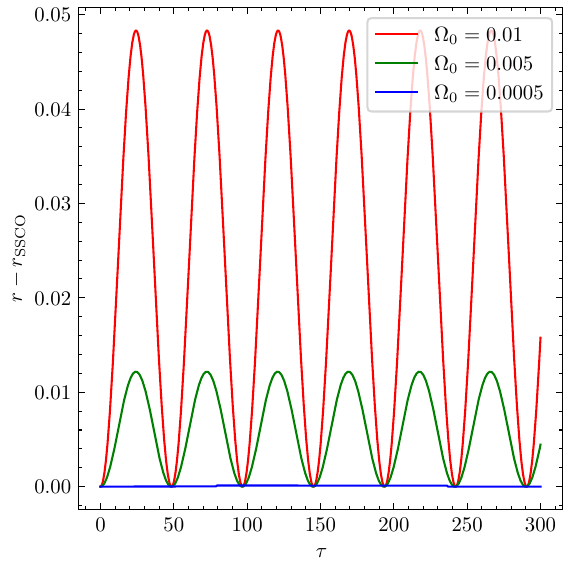}\label{fig64}
}
\caption{ (a) The particle trajectory diagram, where the dashed reference circle has a radius of $4.065175$, the solid orange circle represents the starting position, the solid square represents the ending position. The initial conditions are: radial coordinate \( r_0 = 4.065175 \), angular velocity \( \Omega_0 \), and zero radial velocity. (b) The radial deviation as a function of proper time.  The solution is from Branch 1, with parameters $f_0=0.9$ and $\lambda=10.0$.}
\label{trajectory2}
\end{figure*}

First, we test the stability of the orbit under angular velocity perturbations. As shown in Fig.~\ref{trajectory2}, we fix the initial orbital radius of the particle at $r_\mathrm{SSTCO}$ with zero radial velocity and apply angular velocity perturbations. The dashed circle represents the SSTCO, and the particles are evolved for 300 proper time units. The results indicate: (1) As shown in Fig.~\ref{trajectory2}a, all particles remain in motion near the SSTCO; (2) As shown in Fig.~\ref{trajectory2}b, the amplitude of orbital deviation gradually decreases as the initial angular velocity decreases.

\begin{figure*}[htb]
\centering
\subcaptionbox{}{
\includegraphics[width=0.45\linewidth]{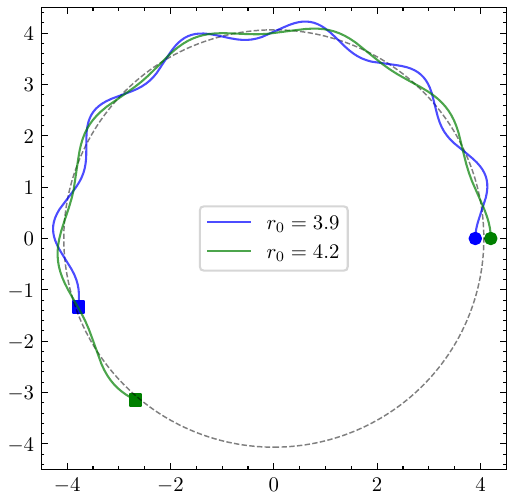}\label{fig61}
}
\subcaptionbox{}{
\includegraphics[width=0.45\linewidth]{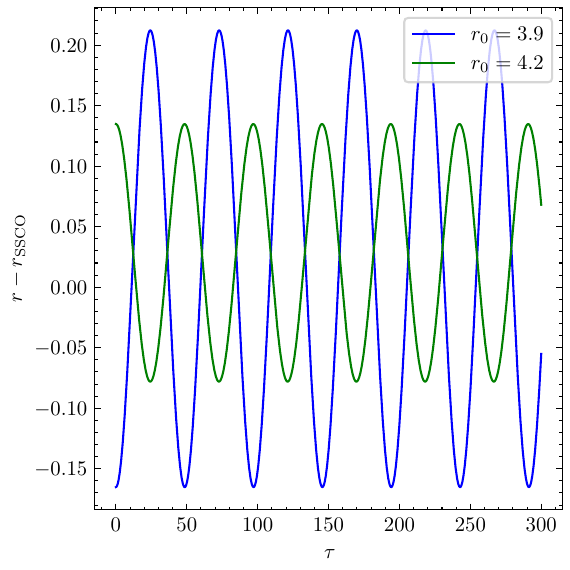}\label{fig62}
}
\caption{(a) The particle trajectory diagram, the dashed reference circle has radius $4.065175$, the solid orange circle represents the starting position, the solid square represents the ending position. The initial conditions are: radial coordinate \( r_0\), angular velocity \( \Omega_0= 0.01 \), and zero radial velocity. (b) The radial deviation as a function of proper time. The solution is from Branch 1, with parameters $f_0=0.9$ and $\lambda=10.0$.}
\label{trajectory1}
\end{figure*}

Second, we test the stability of the orbit under radial perturbations. As depicted in Fig.~\ref{trajectory1}, we fix the initial angular velocity at $\Omega_0 = 0.01$ with zero radial velocity and set initial radii of $r_0 = 3.9$ and $r_0 = 4.2$ as radial perturbations, evolving again over 300 proper time units. Fig.~\ref{trajectory1}a reveals that the particles undergo bounded oscillations around the SSTCO, with radial deviations shown in Fig.~\ref{trajectory1}b, where the maximum deviation remains constant.

Combining the results from Figs.~\ref{trajectory2} and~\ref{trajectory1}, it is evident that whether perturbations are applied to the angular velocity or the radial position, the particle's trajectory remains confined near the SSTCO. This confirms that the orbit with radius $r_{\mathrm{SSTCO}}$ indeed corresponds to the SSTCO.

	Moreover, when $r<r_\mathrm{SSTCO}$, $f'(r)<0$, which is not allowed because circular orbits require $f'(r) \ge 0$ (see Eq.~\eqref{eqO}). Therefore, $r_\mathrm{SSTCO}$ represents the innermost stable circular orbit (ISCO) permitted, i.e., 
\begin{equation}
r_\mathrm{ISCO} = r_\mathrm{SSTCO}. 
\end{equation}

\section{Aschenbach effect}\label{aschenbach}

In a spherically symmetric static spacetime, the orbital velocity $v^{(\phi)} $ of a massive particle moving along a circular orbit, as measured by a ZAMO, can be derived by simplifying the corresponding expression in Kerr spacetime \cite{MuellerCamenzind2004,Aschenbach2004}:
\[
v^{(\phi)}= \sqrt{\frac{- r g_{tt,r}}{2 g_{tt}^2}} = \frac{\sqrt{r f'(r)/2}}{f(r)} .
\]

\begin{figure*}[t]
\centering
\subcaptionbox{$v^{(\phi)}$ in Branch 1}{
\includegraphics[width=0.45\linewidth]{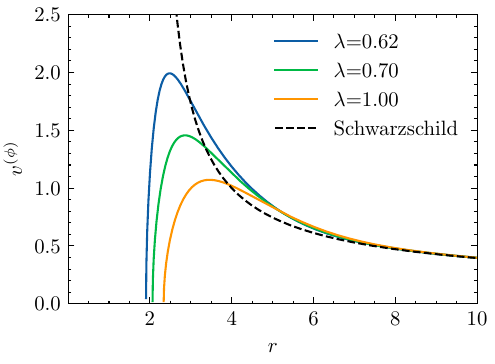}
}
\subcaptionbox{$\Omega$ in Branch 1}{
\includegraphics[width=0.45\linewidth]{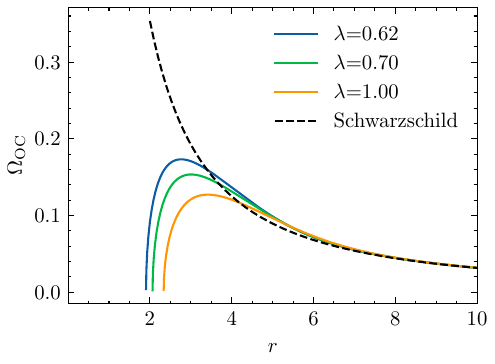}
}

\subcaptionbox{$v^{(\phi)}$ in Branch 2}{
\includegraphics[width=0.45\linewidth]{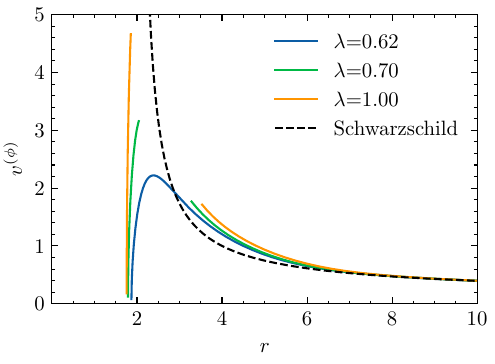}
}
\subcaptionbox{$\Omega$ in Branch 2}{
\includegraphics[width=0.45\linewidth]{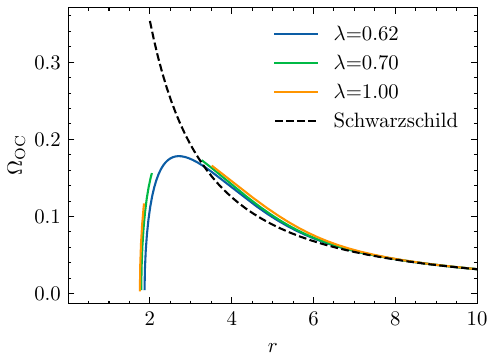}
}
\caption{The distribution of the orbital velocity $v^{(\phi)}$ of massive particles relative to ZAMO, and the distribution of the angular velocity $\Omega_{\mathrm{TCO}}$ of massive particles relative to a distant observer. The first row corresponds to Branch 1, and the second row corresponds to Branch 2, with the parameter $f_0=0.9$. The dashed line represents the Schwarzschild solution.}
\label{v_phi}
\end{figure*}

In the Schwarzschild case, the orbital velocity $v^{(\phi)}$ monotonically increases as $r$ decreases. For cubic gravity, far from the center, $v^{(\phi)}$ behaves like the Schwarzschild case, increasing as $r$ decreases. However, near the center, $v^{(\phi)}$ decreases as $r$ diminishes, exhibiting the typical Aschenbach effect, as shown in Figs.~\ref{v_phi}a and~\ref{v_phi}c. At the same time, Figs.~\ref{v_phi}b and ~\ref{v_phi}d show that the variation of the circular orbital angular velocity $\Omega_{\mathrm{TCO}}$ with $r$ is completely consistent with the variation of $v^{(\phi)}$. This further supports the viewpoint that in a spherically symmetric static spacetime, the non-monotonic behavior of orbital angular velocity with respect to the radial coordinate can serve as an effective criterion for the Aschenbach effect \cite{WeiLiu2023,YerraMukherjiBhamidipati2025,AfsharSadeghi2025}.
Furthermore, in Branch 2, the presence of regions with $f'(r) < 0$ and $D(r) < 0$ (as shown in Fig.~\ref{solDf}) leads to discontinuities in the curves of $v^{(\phi)}$ and $\Omega_{\mathrm{TCO}}$, as illustrated in Figs.~\ref{v_phi}c and~\ref{v_phi}d.

\subsection{Oscillations}

If a particle undergoing circular motion in the equatorial plane is perturbed, it will undergo small oscillations in both the radial and vertical directions \cite{NowakLehr1998}.
In addition to the non-monotonic behavior of $v^{(\phi)}$ and $\Omega_{\mathrm{TCO}}$, the ECG horizonless spacetimes exhibit a rich structure in the epicyclic frequency spectrum. For a static spherically symmetric metric, the vertical epicyclic frequency is degenerate with the orbital frequency (Kepler frequency) for a distant observer \cite{NowakLehr1998,KatoFukueMineshige2008,AbramowiczFragile2013}:
\begin{equation}
\Omega_\theta (r) = \Omega_{\mathrm{TCO}}(r) = \sqrt{\frac{f'(r)}{2r}},
\end{equation}
while the radial epicyclic frequency reads \cite{AbbasRehmanUsama2023}
\begin{equation}\label{Omg3}
\begin{aligned}
\Omega_{r}^{2}(r)=&\frac{1}{2 E_{\mathrm{TCO}}^{2}r^{4}}\left[\left(\left(r^{2}+L_{\mathrm{TCO}}^{2}\right) 3f(r)-2E_{\mathrm{TCO}}^{2}r^{2}\right) r^{2}f(r)f^{\prime\prime}(r)\right.\\
& +2r^{2}\left(\left(r^{2}+L_{\mathrm{TCO}}^{2}\right) 3f(r)-E_{\mathrm{TCO}}^{2}r^{2}\right) f^{\prime 2}(r)\\
&\left.-6 L_{\mathrm{TCO}}^{2}f^{2}(r)(2rf^{\prime}(r)-f(r))\right] .
\end{aligned}
\end{equation}
Substituting $L_{\mathrm{TCO}}$ and $E_{\mathrm{TCO}}$ from Eq. \eqref{L} into the above equation \eqref{Omg3}, we can obtain:
\begin{equation}
\Omega_r^2(r) =  \frac{1}{2} f(r) f''(r) - f'^2(r) + \frac{3}{2}\frac{f(r) f'(r)}{r}.
\label{eq:OmR}
\end{equation}
For a Schwarzschild black hole, Eq.~\eqref{eq:OmR} gives $\Omega_r^2 = \Omega_\theta^2 (1-6M/r)$, which vanishes at the ISCO $r=6M$.

\begin{figure*}[t]
\centering
\subcaptionbox{Branch 1}{
\includegraphics[width=0.45\linewidth]{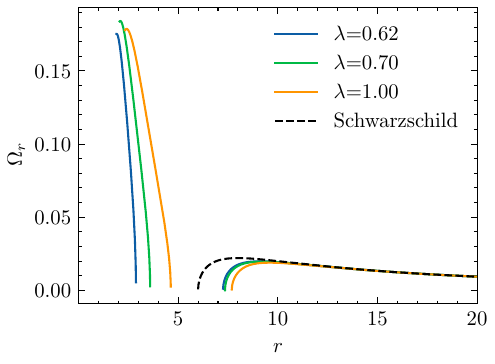}
}
\subcaptionbox{Branch 2}{
\includegraphics[width=0.45\linewidth]{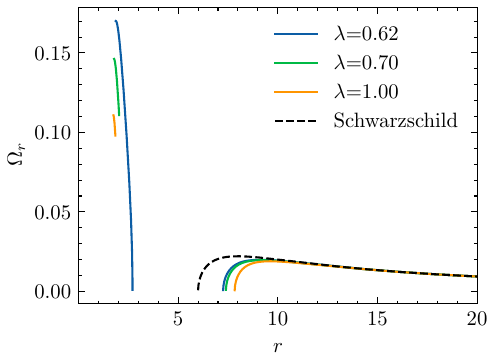}
}

\caption{ The plot of $\Omega_r$ as a function of $r$. Parameter value of $f_0$ is $0.9$. The dashed line represents the Schwarzschild solution.
}
\label{Omega_r}
\end{figure*}

As shown in Fig.~\ref{Omega_r}, the distributions of $\Omega_r$ in both solution branches are discontinuous. This is because the distribution of stable circular orbits is not continuous. Moreover, particles on continuous circular orbits closer to the gravitational center can attain larger $\Omega_r$ values than those on orbits farther from the center.

\begin{figure*}[t]
\centering
\subcaptionbox{Branch 1}{
\includegraphics[width=0.45\linewidth]{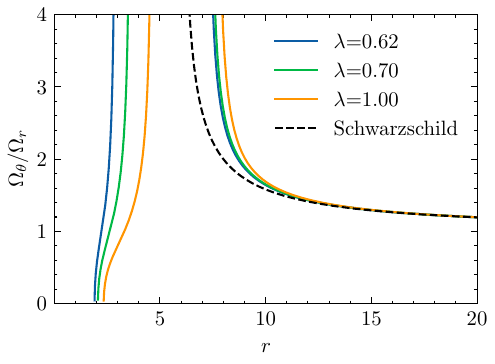}
}
\subcaptionbox{Branch 2}{
\includegraphics[width=0.45\linewidth]{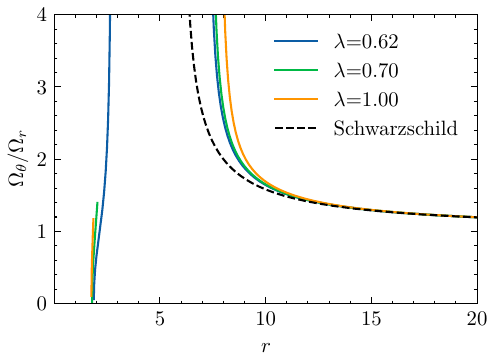}
}

\caption{The ratio of $\Omega_\theta$ to $\Omega_r$. Parameter value of $f_0$ is $0.9$. The dashed line represents the Schwarzschild solution. 
}
\label{Omega_rt}
\end{figure*}

We further illustrate the radial distribution of \(\Omega_\theta / \Omega_r\) in Figure~\ref{Omega_rt}. Due to the constraints imposed by the distributions of \(\Omega_\theta\) (\(\Omega_{\mathrm{TCO}}\)) and \(\Omega_r\), this ratio is also discontinuous in space. Notably, the resonant orbits of interest---those with \(\Omega_\theta : \Omega_r = 3:1\) or \(3:2\)~\cite{Aschenbach2004}---each admit two solutions in Branch~1. However, in Branch~2, such double solutions appear only when \(\lambda\) is near its minimum. In contrast, the Schwarzschild solution yields only one solution for each of the \(3:1\) and \(3:2\) resonant orbits.

\subsection{Specific energy}\label{energy}

If a massive particle undergoes circular motion, equation \eqref{geodesic} yields
\begin{equation}\label{tdot2}
\dot{t} = \frac{1}{\sqrt{-(g_{tt} +  \Omega^2 g_{\phi\phi})}}.
\end{equation}
Substituting the above into \eqref{tdot1} gives the specific energy $E$ satisfying the following equation
\begin{equation}\label{E}
E = -\frac{g_{tt}}{\sqrt{-(g_{tt} +  \Omega^2 g_{\phi\phi}) }}.
\end{equation}

\begin{figure*}[t]
\centering
\subcaptionbox{Branch 1 (full)}{
\includegraphics[width=0.45\linewidth]{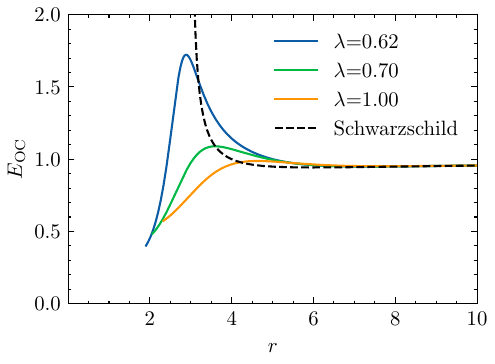}
}
\subcaptionbox{Branch 1 (stable)}{
\includegraphics[width=0.45\linewidth]{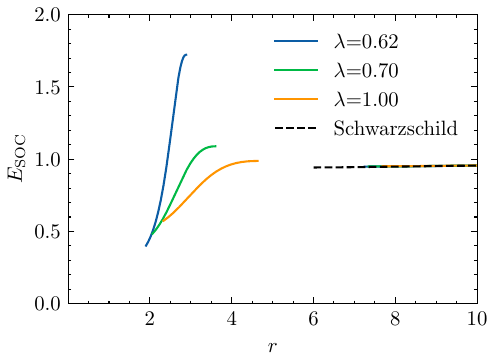}
}

\subcaptionbox{Branch 2 (full)}{
\includegraphics[width=0.45\linewidth]{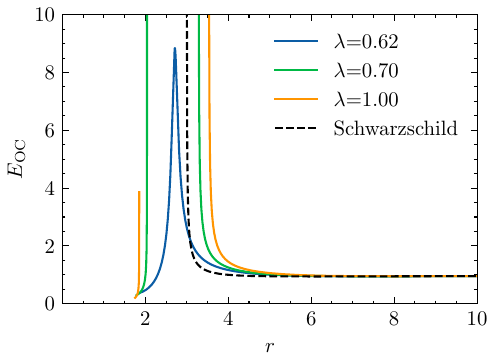}
}
\subcaptionbox{Branch 2 (stable)}{
\includegraphics[width=0.45\linewidth]{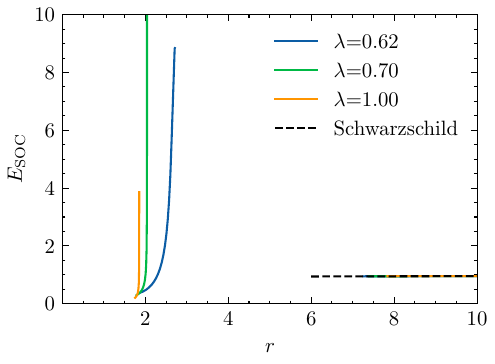}
}
\caption{Distribution of the specific energy of particles in circular orbits, where (a) and (c) correspond to all circular orbits, while (b) and (d) correspond to stable circular orbits. Parameter value of $f_0$ is $0.9$. The dashed line represents the Schwarzschild solution.
}
\label{Eco}
\end{figure*}

In addition to $v^{(\phi)}$ and $\Omega_{\mathrm{TCO}}$, the specific energy $E$ of a massive particle on a circular orbit also exhibits a non-monotonic variation with the orbital radius, as shown in Fig.~\ref{Eco}. Here, Fig.~\ref{Eco}a shows the case of Branch 1, where $E$ remains finite across the entire region. However, for the case of  Branch 2  depicted in Fig.~\ref{Eco}c, $E$ diverges at certain locations, similar to the case of a Schwarzschild black hole, where the divergence points correspond to the position of the photon sphere. 

For null geodesics, the effective potential takes the form:
\begin{equation}\label{Veff2}
V_\mathrm{eff, photon}(r) =  \frac{ f(r)}{r^2},
\end{equation}
whose derivative yields the photon sphere condition
\begin{equation}\label{Veff_derivative}
V'_\mathrm{eff, photon}(r) =  \frac{ rf'(r) - 2 f(r)}{r^3}= -\frac{2 D(r)}{r^3}=0,
\end{equation}
i.e., the location where $D(r)=0$. As indicated by Eq. \eqref{L}, the specific energy $E$ of the particle tends to diverge at the photon sphere radius where $D(r)=0$.

Figures~\ref{Eco}b and~\ref{Eco}d illustrate the distribution of the specific energy $E$ along stable circular orbits. Notably, the regions hosting stable circular orbits are split in both solution branches. The inner stable region extends deep into the gravitational potential.

For Branch 2, within the inner stable region, the specific energy $E$ of massive particles can exceed $1.5$. As the orbital radius decreases, $E$ monotonically drops to a minimum of approximately $0.5$. Consequently, when a test particle migrates from the outer edge to the inner edge (i.e., the ISCO) of this region, the change in specific energy satisfies $\Delta E > 1$. This indicates that during this inward transition, the system releases energy exceeding the particle's rest-mass energy.

For Branch 1, a similar phenomenon can also occur, but it requires $\lambda$ to be close to its minimum value.

\section{Conclusions and discussions}\label{conclusions}

Within the framework of Einsteinian Cubic Gravity (ECG), we investigated the structure of timelike geodesics around a class of static, spherically symmetric, horizonless compact objects. The study revealed that in such purely gravitational, electrically neutral spacetimes, there exist static stable circular orbits. These orbits not only allow test particles to remain static relative to an observer at infinity but also correspond to the innermost stable circular orbit (ISCO) within this spacetime. 
If such static orbits exist in an accretion disk, they could, in principle, produce a stationary feature in the image with reduced Doppler shift compared to ordinary circular orbits. This might serve as a potential criterion, although detailed radiative transfer simulations are needed to confirm.

Through numerical analysis, we found that at larger radii, where the spacetime resembles the Schwarzschild solution, the angular velocity $\Omega_{\mathrm{TCO}}$ and orbital velocity $v^{(\phi)}$ monotonically increase as the radius decreases. However, as the orbit approaches the gravitational center, both velocities exhibit non-monotonic behavior—first increasing and then decreasing until they drop to zero at the static orbit, which is a typical manifestation of the Aschenbach effect. This effect has been observed in both branches of the ECG solution. Additionally, in Branch 2, the region permitting circular orbits becomes  discontinuous due to the emergence of regions where $D(r)<0$ and $f'(r)<0$.  Consequently,  the distribution of stable circular orbits splits into two disconnected domains in both Branches.

Furthermore, our analysis of the specific energy reveals that, for Branch~2, a test particle migrating from the outer edge to the inner edge (ISCO) of the inner stable region can release energy exceeding its rest mass. 
While this phenomenon persists in Branch~1, it requires $\lambda$ to be near its minimum value.

\acknowledgments
This work is supported by the National Natural Science Foundation of China (Grant Nos. 12275110, 12247101) and the National Key Research and Development Program of China (Grant Nos. 2022YFC2204101, 2020YFC2201503).

\bibliographystyle{JHEP}  
\bibliography{library.bib} 

@article{AbbasRehmanUsama2023,
  title = {Accretion Disc around Black Hole in {{Einstein-SU}}({{N}}) Non-Linear Sigma Model},
  author = {Abbas, G. and Rehman, Hamza and Usama, M. and Zhu, Tao},
  year = 2023,
  month = may,
  journal = {Eur. Phys. J. C},
  volume = {83},
  number = {5},
  eprint = {2303.02625},
  primaryclass = {astro-ph.HE},
  pages = {422},
  issn = {1434-6052},
  doi = {10.1140/epjc/s10052-023-11600-0},
  urldate = {2026-06-21},
  archiveprefix = {arXiv},
  langid = {english}
}

@article{AbbottAbbottAbbott2016,
  title = {Observation of {{Gravitational Waves}} from a {{Binary Black Hole Merger}}},
  author = {Abbott, B.P. and Abbott, R. and Abbott, T.D. and others},
  year = 2016,
  month = feb,
  journal = {Phys. Rev. Lett.},
  volume = {116},
  number = {6},
  eprint = {1602.03837},
  primaryclass = {gr-qc},
  pages = {061102},
  publisher = {American Physical Society},
  doi = {10.1103/PhysRevLett.116.061102},
  urldate = {2024-12-09},
  archiveprefix = {arXiv},
  collaboration = {LIGO Scientific, Virgo}
}

@article{AbramowiczFragile2013,
  title = {Foundations of {{Black Hole Accretion Disk Theory}}},
  author = {Abramowicz, Marek A. and Fragile, P. Chris},
  year = 2013,
  journal = {Living Rev. Rel.},
  volume = {16},
  number = {1},
  eprint = {1104.5499},
  primaryclass = {astro-ph.HE},
  pages = {1},
  issn = {2367-3613, 1433-8351},
  doi = {10.12942/lrr-2013-1},
  urldate = {2025-04-03},
  archiveprefix = {arXiv}
}

@article{AdairBuenoCano2020,
  title = {Slowly Rotating Black Holes in {{Einsteinian}} Cubic Gravity},
  author = {Adair, Connor and Bueno, Pablo and Cano, Pablo A. and Hennigar, Robie A. and Mann, Robert B.},
  year = 2020,
  month = oct,
  journal = {Phys. Rev. D},
  volume = {102},
  number = {8},
  eprint = {2004.09598},
  primaryclass = {gr-qc},
  pages = {084001},
  issn = {2470-0010, 2470-0029},
  doi = {10.1103/PhysRevD.102.084001},
  urldate = {2025-05-18},
  archiveprefix = {arXiv}
}

@article{AfsharSadeghi2025,
  title = {Mechanisms behind the Aschenbach Effect in Non-Rotating Black Hole Spacetime},
  author = {Afshar, Mohammad Ali S. and Sadeghi, Jafar},
  year = 2025,
  month = feb,
  journal = {Ann. Phys},
  volume = {474},
  eprint = {2412.06357},
  primaryclass = {gr-qc},
  pages = {169953},
  doi = {10.1016/j.aop.2025.169953},
  archiveprefix = {arXiv},
  langid = {english}
}

@article{AkiyamaAlberdiAlef2019e,
  title = {First {{M87 Event Horizon Telescope Results}}. {{I}}. {{The Shadow}} of the {{Supermassive Black Hole}}},
  author = {Akiyama, Kazunori and Alberdi, Antxon and Alef, Walter and others},
  year = 2019,
  month = apr,
  journal = {Astrophys. J. Lett.},
  volume = {875},
  number = {1},
  eprint = {1906.11238},
  primaryclass = {astro-ph.GA},
  pages = {L1},
  issn = {2041-8205, 2041-8213},
  doi = {10.3847/2041-8213/ab0ec7},
  urldate = {2024-12-09},
  archiveprefix = {arXiv},
  collaboration = {Event Horizon Telescope},
  langid = {american}
}

@article{AkiyamaAlberdiAlef2022e,
  title = {First {{Sagittarius A}}* {{Event Horizon Telescope Results}}. {{I}}. {{The Shadow}} of the {{Supermassive Black Hole}} in the {{Center}} of the {{Milky Way}}},
  author = {Akiyama, Kazunori and Alberdi, Antxon and Alef, Walter and others},
  year = 2022,
  month = may,
  journal = {Astrophys. J. Lett.},
  volume = {930},
  number = {2},
  eprint = {2311.08680},
  primaryclass = {astro-ph.HE},
  pages = {L12},
  issn = {2041-8205, 2041-8213},
  doi = {10.3847/2041-8213/ac6674},
  urldate = {2024-12-09},
  archiveprefix = {arXiv},
  collaboration = {Event Horizon Telescope}
}

@article{AlnaesLoggOlgaard2014,
  title = {Unified {{Form Language}}: {{A}} Domain-Specific Language for Weak Formulations of Partial Differential Equations},
  author = {Alnaes, Martin S. and Logg, Anders and {\O}lgaard, Kristian B. and Rognes, Marie E. and Wells, Garth N.},
  year = 2014,
  journal = {ACM Trans. Math. Softw.},
  volume = {40},
  doi = {10.1145/2566630}
}

@article{Aschenbach2004,
  title = {Measuring Mass and Angular Momentum of Black Holes with High-Frequency Quasi-Periodic Oscillations},
  author = {Aschenbach, B.},
  year = 2004,
  journal = {Astron. Astrophys.},
  volume = {425},
  number = {3},
  eprint = {astro-ph/0406545},
  pages = {1075--1082},
  publisher = {EDP Sciences},
  issn = {0004-6361, 1432-0746},
  doi = {10.1051/0004-6361:20041412},
  urldate = {2025-09-27},
  archiveprefix = {arXiv},
  copyright = {\copyright{} ESO, 2004},
  langid = {english}
}

@inproceedings{Aschenbach2004a,
  title = {Mass and Angular Momentum of {{Sgr A}}*},
  booktitle = {Conf. {{Grow}}. {{Black Holes Accretion Cosmol}}. {{Context}}},
  author = {Aschenbach, Bernd},
  year = 2004,
  month = oct,
  eprint = {astro-ph/0410328},
  doi = {10.1007/11403913_54},
  archiveprefix = {arXiv},
  langid = {english}
}

@article{Aschenbach2006,
  title = {Mass and {{Angular Momentum}} of {{Black Holes}}: {{An Overlooked Effect}} of {{General Relativity Applied}} to the {{Galactic Center Black Hole Sgr A}}*},
  shorttitle = {Mass and {{Angular Momentum}} of {{Black Holes}}},
  author = {Aschenbach, B.},
  year = 2006,
  journal = {Chin. J. Astron. Astrophys.},
  volume = {6},
  number = {Suppl.1},
  eprint = {astro-ph/0603193},
  pages = {1221},
  issn = {1009-9271},
  doi = {10.1088/1009-9271/6/S1/28},
  urldate = {2025-10-09},
  archiveprefix = {arXiv},
  langid = {english}
}

@article{Bakhtiarizadeh2022,
  title = {Charged Rotating Black Strings in {{Einsteinian}} Cubic Gravity},
  author = {Bakhtiarizadeh, Hamid R.},
  year = 2022,
  month = mar,
  journal = {Phys. Rev. D},
  volume = {105},
  number = {6},
  eprint = {2111.02663},
  primaryclass = {gr-qc},
  pages = {064037},
  doi = {10.1103/PhysRevD.105.064037},
  archiveprefix = {arXiv},
  langid = {english}
}

@article{BakhtiarizadehGolchin2022,
  title = {Rotating Black Strings in {{Einsteinian}} Cubic Gravity with Born-Infeld Electrodynamics},
  author = {Bakhtiarizadeh, Hamid Reza and Golchin, Hanif},
  year = 2022,
  month = nov,
  journal = {Iran. J. Phys. Res.},
  volume = {22},
  number = {3},
  pages = {663--670},
  doi = {10.47176/ijpr.22.3.51460},
  langid = {english}
}

@misc{BarattaDeanDokken2023,
  title = {{{DOLFINx}}: {{The}} next Generation {{FEniCS}} Problem Solving Environment},
  shorttitle = {{{DOLFINx}}},
  author = {Baratta, Igor A. and Dean, Joseph P. and Dokken, J{\o}rgen S. and Habera, Michal and Hale, Jack S. and Richardson, Chris N. and Rognes, Marie E. and Scroggs, Matthew W. and Sime, Nathan and Wells, Garth N.},
  year = 2023,
  month = dec,
  publisher = {Zenodo},
  doi = {10.5281/zenodo.10447666},
  urldate = {2024-08-02},
  archiveprefix = {Zenodo},
  langid = {english}
}

@article{BuenoCano2016,
  title = {Einsteinian Cubic Gravity},
  author = {Bueno, Pablo and Cano, Pablo A.},
  year = 2016,
  month = nov,
  journal = {Phys. Rev. D},
  volume = {94},
  number = {10},
  eprint = {1607.06463},
  primaryclass = {hep-th},
  pages = {104005},
  issn = {2470-0010, 2470-0029},
  doi = {10.1103/PhysRevD.94.104005},
  urldate = {2025-08-07},
  archiveprefix = {arXiv}
}

@article{BuenoCano2016a,
  title = {Four-Dimensional Black Holes in {{Einsteinian}} Cubic Gravity},
  author = {Bueno, Pablo and Cano, Pablo A.},
  year = 2016,
  month = dec,
  journal = {Phys. Rev. D},
  volume = {94},
  number = {12},
  eprint = {1610.08019},
  primaryclass = {hep-th},
  pages = {124051},
  publisher = {American Physical Society},
  doi = {10.1103/PhysRevD.94.124051},
  urldate = {2025-05-18},
  archiveprefix = {arXiv}
}

@article{BuenoCanoMoreno2019,
  title = {All Higher-Curvature Gravities as {{Generalized}} Quasi-Topological Gravities},
  author = {Bueno, Pablo and Cano, Pablo A. and Moreno, Javier and Murcia, {\'A}ngel},
  year = 2019,
  month = nov,
  journal = {J. High Energy Phys.},
  volume = {11},
  number = {11},
  eprint = {1906.00987},
  primaryclass = {hep-th},
  pages = {062},
  issn = {1029-8479},
  doi = {10.1007/JHEP11(2019)062},
  urldate = {2025-12-01},
  archiveprefix = {arXiv},
  langid = {english}
}

@article{BurgerEmondMoynihan2020,
  title = {Rotating Black Holes in Cubic Gravity},
  author = {Burger, Daniel J. and Emond, William T. and Moynihan, Nathan},
  year = 2020,
  month = apr,
  journal = {Phys. Rev. D},
  volume = {101},
  number = {8},
  eprint = {1910.11618},
  primaryclass = {hep-th},
  pages = {084009},
  doi = {10.1103/PhysRevD.101.084009},
  archiveprefix = {arXiv},
  langid = {english}
}

@article{CanoPereniguez2020,
  title = {Extremal Rotating Black Holes in {{Einsteinian}} Cubic Gravity},
  author = {Cano, Pablo A. and Pere{\~n}iguez, David},
  year = 2020,
  month = feb,
  journal = {Phys. Rev. D},
  volume = {101},
  number = {4},
  eprint = {1910.10721},
  primaryclass = {hep-th},
  pages = {044016},
  doi = {10.1103/PhysRevD.101.044016},
  archiveprefix = {arXiv},
  langid = {english}
}

@article{CisternaGrandiOliva2020,
  title = {On Four-Dimensional {{Einsteinian}} Gravity, Quasitopological Gravity, Cosmology and Black Holes},
  author = {Cisterna, Adolfo and Grandi, Nicol{\'a}s and Oliva, Julio},
  year = 2020,
  month = jun,
  journal = {Phys. Lett. B},
  volume = {805},
  eprint = {1811.06523},
  primaryclass = {hep-th},
  pages = {135435},
  issn = {0370-2693},
  doi = {10.1016/j.physletb.2020.135435},
  urldate = {2025-12-01},
  archiveprefix = {arXiv}
}

@article{CollodelKleihausKunz2018,
  title = {Static Orbits in Rotating Spacetimes},
  author = {Collodel, Lucas G. and Kleihaus, Burkhard and Kunz, Jutta},
  year = 2018,
  month = may,
  journal = {Phys. Rev. Lett.},
  volume = {120},
  number = {20},
  eprint = {1711.05191},
  primaryclass = {gr-qc},
  pages = {201103},
  publisher = {American Physical Society (APS)},
  issn = {0031-9007, 1079-7114},
  doi = {10.1103/PhysRevLett.120.201103},
  urldate = {2025-07-24},
  archiveprefix = {arXiv},
  copyright = {https://link.aps.org/licenses/aps-default-license},
  langid = {english}
}

@article{Dengiz2025,
  title = {Cosmological {{Weyl-Einsteinian-cubic}} Gravity as a Gauge Theory of Gravity},
  author = {Dengiz, Suat},
  year = 2025,
  month = sep,
  journal = {Eur. Phys. J. C},
  volume = {85},
  number = {9},
  eprint = {2408.04282},
  primaryclass = {hep-th},
  pages = {928},
  doi = {10.1140/epjc/s10052-025-14647-3},
  archiveprefix = {arXiv},
  langid = {english}
}

@article{EricesPapantonopoulosSaridakis2019,
  title = {Cosmology in Cubic and f({{P}}) Gravity},
  author = {Erices, Cristian and Papantonopoulos, Eleftherios and Saridakis, Emmanuel N.},
  year = 2019,
  month = jun,
  journal = {Phys. Rev. D},
  volume = {99},
  number = {12},
  eprint = {1903.11128},
  primaryclass = {gr-qc},
  pages = {123527},
  doi = {10.1103/PhysRevD.99.123527},
  archiveprefix = {arXiv},
  langid = {english}
}

@article{FrassinoRocha2020,
  title = {Charged Black Holes in {{Einsteinian}} Cubic Gravity and Nonuniqueness},
  author = {Frassino, Antonia Micol and Rocha, Jorge V.},
  year = 2020,
  month = jul,
  journal = {Phys. Rev. D},
  volume = {102},
  number = {2},
  eprint = {2002.04071},
  primaryclass = {hep-th},
  pages = {024035},
  doi = {10.1103/PhysRevD.102.024035},
  archiveprefix = {arXiv},
  langid = {english}
}

@article{HennigarMann2017,
  title = {Black Holes in {{Einsteinian}} Cubic Gravity},
  author = {Hennigar, Robie A. and Mann, Robert B.},
  year = 2017,
  month = mar,
  journal = {Phys. Rev. D},
  volume = {95},
  number = {6},
  eprint = {1610.06675},
  primaryclass = {hep-th},
  pages = {064055},
  publisher = {American Physical Society},
  doi = {10.1103/PhysRevD.95.064055},
  urldate = {2025-05-18},
  archiveprefix = {arXiv},
  langid = {american}
}

@article{HussainMustafa2022,
  title = {Traversable Wormholes in {{Einsteinian-cubic-gravity}} with Hybrid Shape Functions},
  author = {Hussain, Ibrar and Mustafa, G.},
  year = 2022,
  month = feb,
  journal = {Int. J. Geom. Meth. Mod. Phys.},
  volume = {19},
  number = {05},
  pages = {2250074},
  doi = {10.1142/S0219887822500748},
  langid = {english}
}

@book{KatoFukueMineshige2008,
  title = {Black-{{Hole Accretion Disks}} : {{Towards}} a {{New Paradigm}}},
  author = {Kato, S. and Fukue, J. and Mineshige, S.},
  year = 2008,
  journal = {Black-Hole Accretion Disks --- Towards a New Paradigm},
  publisher = {Kyoto University Press},
  address = {Kyoto, Japan},
  urldate = {2026-06-26},
  isbn = {978-4-87698-740-5}
}

@article{KhodagholizadehPerlickVahedi2020,
  title = {Aschenbach Effect for Spinning Particles in Kerr Spacetime},
  author = {Khodagholizadeh, Jafar and Perlick, Volker and Vahedi, Ali},
  year = 2020,
  month = jul,
  journal = {Phys. Rev. D},
  volume = {102},
  number = {2},
  eprint = {2002.04701},
  primaryclass = {gr-qc},
  pages = {24021},
  doi = {10.1103/PhysRevD.102.024021},
  archiveprefix = {arXiv},
  langid = {english}
}

@article{KordZangenehKazemi2020,
  title = {Topological Born--Infeld Charged Black Holes in {{Einsteinian}} Cubic Gravity},
  author = {Kord Zangeneh, M. and Kazemi, A.},
  year = 2020,
  month = aug,
  journal = {Eur. Phys. J. C},
  volume = {80},
  number = {8},
  eprint = {2003.04458},
  primaryclass = {hep-th},
  pages = {794},
  doi = {10.1140/epjc/s10052-020-8394-8},
  archiveprefix = {arXiv},
  langid = {english}
}

@article{LessaSilva2023,
  title = {Regular Black Holes in {{Einstein}} Cubic Gravity},
  author = {Lessa, L.A. and Silva, J.E.G.},
  year = 2023,
  month = may,
  eprint = {2305.18254},
  primaryclass = {gr-qc},
  publisher = {arXiv},
  archiveprefix = {arXiv},
  langid = {english}
}

@article{LuYangMann2025,
  title = {Existence of Vacuum Wormholes in {{Einsteinian}} Cubic Gravity},
  author = {Lu, Mengqi and Yang, Jiayue and Mann, Robert B.},
  year = 2025,
  month = mar,
  journal = {JHEP},
  volume = {3},
  eprint = {2410.13996},
  primaryclass = {gr-qc},
  pages = {73},
  doi = {10.1007/JHEP03(2025)073},
  archiveprefix = {arXiv}
}

@article{MehdizadehZiaie2019,
  title = {Traversable Wormholes in {{Einsteinian}} Cubic Gravity},
  author = {Mehdizadeh, Mohammad Reza and Ziaie, Amir Hadi},
  year = 2019,
  month = oct,
  journal = {Mod. Phys. Lett. A},
  volume = {35},
  number = {06},
  eprint = {1903.10907},
  primaryclass = {gr-qc},
  pages = {2050017},
  doi = {10.1142/S0217732320500170},
  archiveprefix = {arXiv},
  langid = {english}
}

@article{MuellerAschenbach2007,
  title = {Non-Monotonic Orbital Velocity Profiles around Rapidly Rotating Kerr-(Anti-)de Sitter Black Holes},
  author = {Mueller, Andreas and Aschenbach, Bernd},
  year = 2007,
  journal = {Cl. Quant Grav},
  volume = {24},
  eprint = {0704.3963},
  primaryclass = {gr-qc},
  pages = {2637--2644},
  doi = {10.1088/0264-9381/24/10/009},
  archiveprefix = {arXiv},
  langid = {english}
}

@article{MuellerCamenzind2004,
  title = {Relativistic Emission Lines from Accreting Black Holes - {{The}} Effect of Disk Truncation on Line Profiles},
  author = {Mueller, Andreas and Camenzind, Max},
  year = 2004,
  journal = {Astron. Astrophys.},
  volume = {413},
  number = {3},
  eprint = {astro-ph/0309832},
  pages = {861--878},
  issn = {0004-6361, 1432-0746},
  doi = {10.1051/0004-6361:20031522},
  urldate = {2025-10-09},
  archiveprefix = {arXiv}
}

@article{MustafaAtamurotovGhosh2023,
  title = {Structural Properties of Generalized Embedded Wormhole Solutions via Dark Matter Halos in {{Einsteinian-cubic-gravity}} with Quasi-Periodic Oscillations},
  author = {Mustafa, G. and Atamurotov, Farruh and Ghosh, Sushant G.},
  year = 2023,
  month = mar,
  journal = {Phys. Dark Univ.},
  volume = {40},
  pages = {101214},
  doi = {10.1016/j.dark.2023.101214},
  langid = {english}
}

@article{MustafaXiaHussain2020,
  title = {Spherically Symmetric Static Wormhole Models in the {{Einsteinian}} Cubic Gravity},
  author = {Mustafa, G. and Xia, Tie-Cheng and Hussain, Ibrar and Shamir, M. Farasat},
  year = 2020,
  month = nov,
  journal = {Int. J. Geom. Meth. Mod. Phys.},
  volume = {17},
  number = {14},
  pages = {2050214},
  doi = {10.1142/S021988782050214X},
  langid = {english}
}

@incollection{NowakLehr1998,
  title = {Stable Oscillations of Black Hole  Accretion Discs},
  booktitle = {Theory of {{Black Hole Accretion Disks}}},
  author = {Nowak, M. A. and Lehr, D. E.},
  editor = {Abramowicz, Marek A. and Bj{\"o}rnsson, Gunnlaugur and Pringle, James E.},
  year = 1998,
  pages = {233},
  publisher = {Cambridge University Press},
  urldate = {2026-06-26},
  isbn = {0-521-62362-6}
}

@article{QuirosGarcia-SalcedoGonzalez2020,
  title = {Global Asymptotic Dynamics of Cosmological {{Einsteinian}} Cubic Gravity},
  author = {Quiros, Israel and {Garc{\'i}a-Salcedo}, Ricardo and Gonzalez, Tame and Mart{\'i}nez, Jorge Luis Morales and Nucamendi, Ulises},
  year = 2020,
  month = aug,
  journal = {Phys. Rev. D},
  volume = {102},
  number = {4},
  eprint = {2003.10516},
  primaryclass = {gr-qc},
  pages = {044018},
  doi = {10.1103/PhysRevD.102.044018},
  archiveprefix = {arXiv},
  langid = {english}
}

@article{SajadiHendi2022,
  title = {Analytically Approximation Solution to Einstein-Cubic Gravity},
  author = {Sajadi, S. N. and Hendi, S. H.},
  year = 2022,
  month = aug,
  journal = {Eur. Phys. J. C},
  volume = {82},
  number = {8},
  eprint = {2207.13435},
  primaryclass = {gr-qc},
  pages = {675},
  doi = {10.1140/epjc/s10052-022-10647-9},
  archiveprefix = {arXiv},
  langid = {english}
}

@article{ScroggsBarattaRichardson2022,
  title = {Basix: A Runtime Finite Element Basis Evaluation Library},
  author = {Scroggs, Matthew W. and Baratta, Igor A. and Richardson, Chris N. and Wells, Garth N.},
  year = 2022,
  journal = {J. Open Source Softw.},
  volume = {7},
  number = {73},
  pages = {3982},
  doi = {10.21105/joss.03982}
}

@article{StuchlikBlaschkeSlany2011,
  title = {Non-Monotonic Keplerian Velocity Profiles around near-Extreme Braneworld Kerr Black Holes},
  author = {Stuchlik, Zdenek and Blaschke, Martin and Slany, Petr},
  year = 2011,
  journal = {Cl. Quant Grav},
  volume = {28},
  eprint = {1108.0191},
  primaryclass = {gr-qc},
  pages = {175002},
  doi = {10.1088/0264-9381/28/17/175002},
  archiveprefix = {arXiv},
  langid = {english}
}

@article{StuchlikSlanyTorok2005,
  title = {Aschenbach Effect: Unexpected Topology Changes in the Motion of Particles and Fluids Orbiting Rapidly Rotating Kerr Black Holes},
  shorttitle = {Aschenbach Effect},
  author = {Stuchl{\'i}k, Zden{\v e}k and Slan{\'y}, Petr and T{\"o}r{\"o}k, Gabriel and Abramowicz, Marek A.},
  year = 2005,
  journal = {Phys. Rev. D},
  volume = {71},
  number = {2},
  eprint = {gr-qc/0411091},
  pages = {24037},
  publisher = {American Physical Society},
  doi = {10.1103/PhysRevD.71.024037},
  urldate = {2025-09-27},
  archiveprefix = {arXiv},
  langid = {english}
}

@article{TeodoroCollodelDoneva2021,
  title = {Polish {{Doughnuts}} around {{Scalarized Kerr Black Holes}}},
  author = {Teodoro, Matheus C. and Collodel, Lucas G. and Doneva, Daniela and Kunz, Jutta and Nedkova, Petya and Yazadjiev, Stoytcho},
  year = 2021,
  month = dec,
  journal = {Phys. Rev. D},
  volume = {104},
  number = {12},
  eprint = {2108.08640},
  primaryclass = {gr-qc},
  pages = {124047},
  doi = {10.1103/PhysRevD.104.124047},
  urldate = {2025-09-05},
  archiveprefix = {arXiv}
}

@article{TeodoroCollodelKunz2021,
  title = {Retrograde {{Polish Doughnuts}} around {{Boson Stars}}},
  author = {Teodoro, Matheus C. and Collodel, Lucas G. and Kunz, Jutta},
  year = 2021,
  month = mar,
  journal = {JCAP},
  volume = {03},
  number = {03},
  eprint = {2011.10288},
  primaryclass = {gr-qc},
  pages = {063},
  issn = {1475-7516},
  doi = {10.1088/1475-7516/2021/03/063},
  urldate = {2025-08-13},
  archiveprefix = {arXiv},
  langid = {american}
}

@article{TursunovStuchlikKolos2016,
  title = {Circular Orbits and Related Quasiharmonic Oscillatory Motion of Charged Particles around Weakly Magnetized Rotating Black Holes},
  author = {Tursunov, Arman and Stuchl{\'i}k, Zden{\v e}k and Kolo{\v s}, Martin},
  year = 2016,
  month = apr,
  journal = {Phys. Rev. D},
  volume = {93},
  number = {8},
  eprint = {1603.07264},
  primaryclass = {gr-qc},
  pages = {084012},
  doi = {10.1103/PhysRevD.93.084012},
  archiveprefix = {arXiv},
  langid = {english}
}

@article{VahediKhodagholizadehTursunov2021,
  title = {Aschenbach Effect for Spinning Particles in Kerr-(a){{dS}} Spacetime},
  author = {Vahedi, Ali and Khodagholizadeh, Jafar and Tursunov, Arman},
  year = 2021,
  month = apr,
  journal = {Eur. Phys. J. C},
  volume = {81},
  number = {4},
  eprint = {2103.14912},
  primaryclass = {gr-qc},
  pages = {280},
  doi = {10.1140/epjc/s10052-021-09081-0},
  archiveprefix = {arXiv},
  langid = {english}
}

@article{Wang2024a,
  title = {Frozen Gravitational Stars in {{Einsteinian}} Cubic Gravity},
  author = {Wang, Yong-Qiang},
  year = 2024,
  month = oct,
  eprint = {2410.04575},
  primaryclass = {gr-qc},
  archiveprefix = {arXiv},
  langid = {american}
}

@article{WeiLiu2023,
  title = {Aschenbach Effect and Circular Orbits in Static and Spherically Symmetric Black Hole Backgrounds},
  author = {Wei, Shao-Wen and Liu, Yu-Xiao},
  year = 2023,
  month = dec,
  journal = {Phys. Dark Univ.},
  volume = {43},
  eprint = {2308.11883},
  primaryclass = {gr-qc},
  pages = {101409},
  issn = {22126864},
  doi = {10.1016/j.dark.2023.101409},
  urldate = {2025-07-30},
  archiveprefix = {arXiv},
  langid = {american}
}

@article{WeiZhangLiu2023,
  title = {Static Spheres around Spherically Symmetric Black Hole Spacetime},
  author = {Wei, Shao-Wen and Zhang, Yu-Peng and Liu, Yu-Xiao and Mann, Robert B.},
  year = 2023,
  month = oct,
  journal = {Phys. Rev. Res.},
  volume = {5},
  number = {4},
  eprint = {2303.06814},
  primaryclass = {gr-qc},
  pages = {043050},
  issn = {2643-1564},
  doi = {10.1103/PhysRevResearch.5.043050},
  urldate = {2025-07-30},
  archiveprefix = {arXiv},
  langid = {american}
}

@article{YerraMukherjiBhamidipati2025,
  title = {Static Spheres and {{Aschenbach}} Effect for Black Holes in Massive Gravity},
  author = {Yerra, Pavan Kumar and Mukherji, Sudipta and Bhamidipati, Chandrasekhar},
  year = 2025,
  month = jun,
  journal = {Phys. Rev. D},
  volume = {111},
  number = {12},
  eprint = {2411.01261},
  primaryclass = {gr-qc},
  pages = {124018},
  doi = {10.1103/lj4b-j3tr},
  archiveprefix = {arXiv}
}

\end{document}